\documentclass[fleqn,usenatbib]{mnras}

\usepackage{newtxtext,newtxmath}

\usepackage[T1]{fontenc}
\usepackage{ae,aecompl}

\usepackage{graphicx}	
\usepackage{amsmath}	
\usepackage{amssymb}	

\title[Close encounters between proto-Mercury and Proto-Venus]{Extreme close encounters between proto-Mercury and proto-Venus in terrestrial planet formation}

\author[Fang \& Deng]{
  Tong Fang, $^{1}$
Hongping Deng,$^{2}$\thanks{E-mail:hd353@cam.ac.uk} \\
$^{1}$ Center of Deep Sea Research, Institute of Oceanology, Chinese Academy of Sciences, Qingdao 266071, China \\
$^{2}$Department of Applied Mathematics and Theoretical Physics, University of Cambridge, Centre for Mathematical Sciences, Wilberforce Road, Cambridge CB3 0WA, UK
}

\pubyear{2018}

\begin{document}
\label{firstpage}
\pagerange{\pageref{firstpage}--\pageref{lastpage}}
\maketitle

\begin{abstract}
  Modern models of terrestrial planet formation require solids depletion interior to 0.5-0.7 au in the planetesimal disk to explain the small mass of Mercury. Earth and Venus analogues emerge after $\sim$100 Myr collisional growth while Mercury form in the diffusive tails of the planetesimal disk. We carried out 250 N-body simulations of planetesimal disks with mass confined to 0.7-1.0 au to study the statistics of close encounters which were recently proposed as an explanation for the high iron mass fraction in Mercury by \citet{Deng2020}. We formed 39 Mercury analogues in total and all proto-Mercury analogues were scattered inward by proto-Venus. Proto-Mercury typically experiences 6 extreme close encounters (closest approach smaller than 6 Venus radii) with Proto-Venus after Proto-Venus acquires 0.7 Venus Mass. At such close separation, the tidal interaction can already affect the orbital motion significantly such that the N-body treatment itself is invalid. More and closer encounters are expected should tidal dissipation of orbital angular momentum accounted. Hybrid N-body hydrodynamic simulations, treating orbital and encounter dynamics self-consistently, are desirable to evaluate the degree of tidal mantle stripping of proto-Mercury. 
\end{abstract}




\section{Introduction}
\label{sec:intro}
The terrestrial planets in our solar system form through collisional growth between planetesimals embedded in a protoplanetary disk \citep{Safronov1969}. The growth of dust to planetesimals under the influence of turbulence in protoplanetary disks remains elusive \citep[see, e.g., review by][]{Testi2014}. The later collisions between planetesimals are readily modeled with N-body simulations \citep{Kokubo1998, Kokubo2000, Chambers1998,Morishima2010}. Earth and Venus analogues naturally emerge in N-body simulations \citep{Obrien2006} while replicating the smaller planets, Mars and Mercury remains challenging \citep{Raymond2009}. Recent model starting from planetary embryos confined between 0.7-1 au successfully produced Mars and Mercury analogues from material gravitationally scattered (diffused) outside the original annulus \citep{Hansen2009}. The initial narrow annulus may be caused by the migration of Jupiter and Saturn in the protoplanetary disk \citep{Walsh2011} or it can be intrinsic to the planetesimal distribution due to dust drift in protoplanetary disks \citep{Drazkowska2016, Raymond2017}. The narrow annulus model and its variations are promising with high success rate in replicating Mars and Mercury \citep{Hansen2009}. Alternatively, an eccentric Jupiter \citep[likely caused by giant planet instability, see][]{Clement2018} can excite the planetesimal disk and form small Mars analogues.  

The formation of Mercury, is arguably the most challenging problem in these N-body simulations \citep{Lykawka2017, Clement2019}. To replicate the small mass of Mercury, an \emph{ad hoc} depletion of solids interior to 0.5–0.7 au is typically invoked \citep[see, e.g.,][]{Chambers2001,Hansen2009,Walsh2011,Clement2018} except in the in situ formation model of \citet{Lykawka2017,Lykawka2019}. The solids may be depleted by a massive body which later moved away, e.g, a super-Earth lost to Sun \citep{Batygin2015} or the Jupiter core migrated outwards \citep{Raymond2016}. The solids deficit can also be caused by the lost of early formed embryos to Sun \citep{Ida2008} or bares the imprint of the fossil silicate snowline \citep{Morbidelli2016}. In these models, Mercury itself or its building blocks must be scattered inward 0.5 au. If the former is the case then violent encounters between proto-Mercury and proto-Venus may be necessary to kick proto-Mercury in. 

\begin{figure*}
\centering
\includegraphics[width=0.9\linewidth]{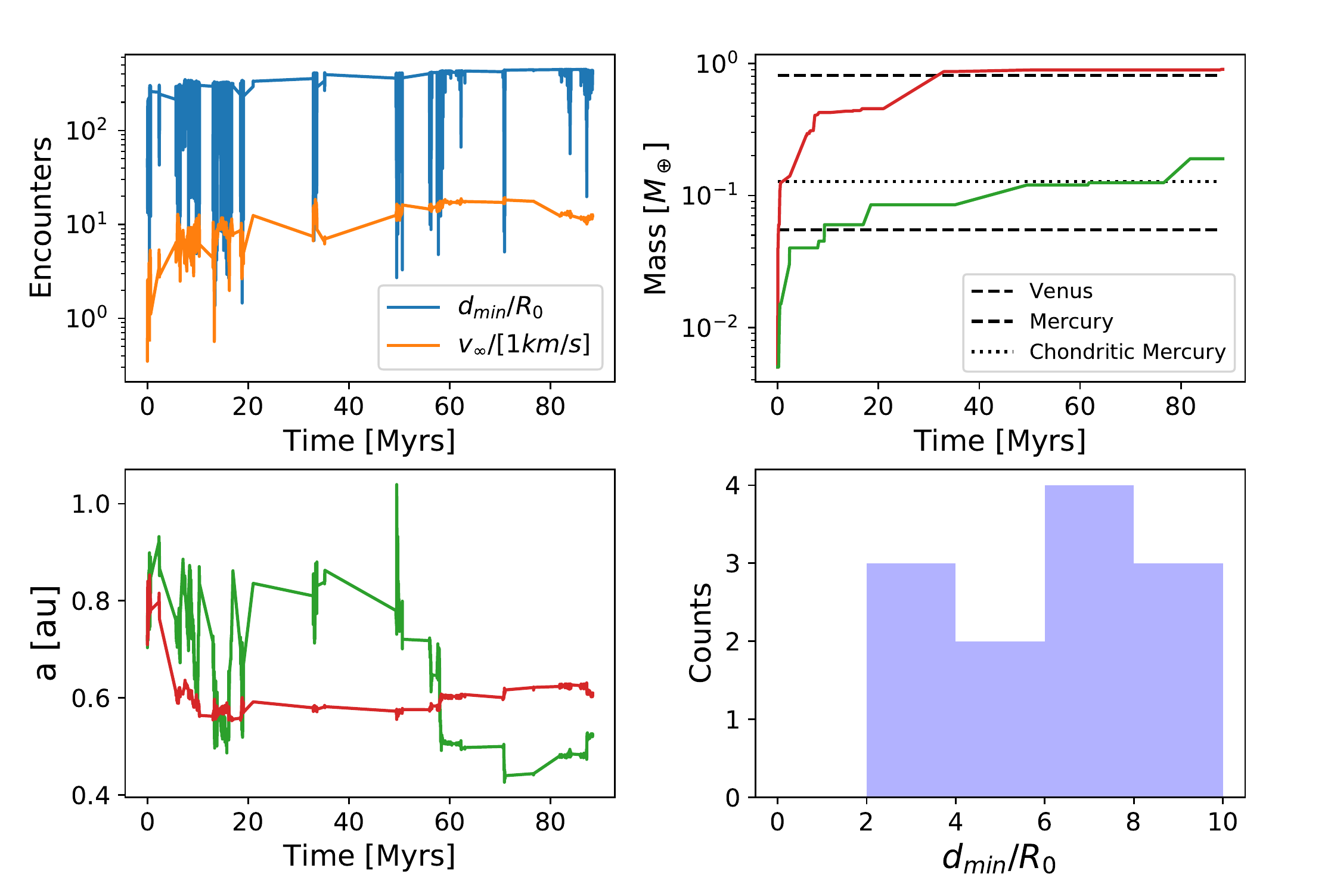}
\caption{Encounters between proto-Mercury and proto-Venus in an exemplary simulation. \emph{Upper left:} the closest approach $d_{min}$ in the units of Venus radius, $R_0$ and the relative velocity at infinity $v_\infty$ for all the close encounter between the formed Mercury and Venus analogues. \emph{Upper right:} mass growth of Mercury and Venus analogues. \emph{Lower left:} semi-major axis of proto-Mercury and proto-Venus at the time of their close encounters. \emph{Lower right:} the number of major encounters with $d_{min}$ smaller than $10R_0$ (see text).\label{fig:eg}.}
\end{figure*}

Mercury also stands out for its anonymously high-iron mass fraction, $70\%$ of the planet mass \citep{Anderson1987, Hauck2013} while all other terrestrial planets has only 30\% iron by mass. This rocky material deficit may be a result of mantle removal due to giant impacts \citep{Benz2007, Asphaug2014, Chau2018} or extreme close encounters with proto-Venus \citep{Deng2020}. The high-temperature giant impact scenario seems at odds with the retainment of moderately volatile elements on the present-day Mercury \citep{Peplowski2011}. The multiple encounters scenario avoids high-temperature collisions but whether such extreme close encounters occur repeatedly remains uncertain.  \cite{Deng2020} shows mantle removal occurs for a non-spinning proto-Mercury passing by proto-Venus within 2 Venus radii. Although a prograde spin enhances mantle removal, at least 4 close encounters are needed to deplete Mercury mantle to present-day level.

Here we use N-body simulations of terrestrial planet formation to find out whether close encounter between proto-Mercury and proto-Venus occurs. We focus on those extreme encounters which can potentially lead to mantle removal of proto-Mercury. We describe our simulations and present our results in section \ref{sec:simulaiton} and section \ref{sec:results}. Discussion follows in section \ref{sec:diss}. We then conclude in section \ref{sec:con}.

\section{Simulations}
\label{sec:simulaiton}

We used the \emph{Mercury6} package to integrate the planetary embryo system assuming perfect merging upon collisions \citep{Chambers1999}. We chose the confined annulus model of \cite{Hansen2009} because of its high success rate in producing Mercury analogues \citep{Clement2019}. This model is chosen also because of its simplicity requiring no extra parameters, unlike for example, parameters for the giant planet migration in the Grand Tack model \citep{Walsh2011}. Initially, the system consists of 400 planetary embryos of equal mass, 0.005 $M_\oplus$ (Earth mass) placed between 0.7-1 au representing a flat surface density profile \citep{Hansen2009}. The density of planetary embryos is assumed to be 4 g/cm$^3$. Jupiter is placed at 5 au with eccentricity $=0.05$. We used a timestep of 4 days \citep{Hansen2009, Lykawka2017} and integrated the system for 200 million years \citep{Clement2019, Kaib2015}.

We carried out 250 simulations starting from different realisations of the above solids distribution \citep{Hansen2009}. The evolution of this system is well documented in \cite{Hansen2009} and here we focus on close encounters between Mercury and Venus analogues. At the end of simulations, the large bodies close to 0.7 au and 1 au are identified as Venus and Earth analogues. We note that we exclude systems left with only one dominant body (typically $>1.33M_\oplus$) and three bodies (typically all larger than $0.33M_\oplus$) in later analysis. We formed 39 Mercury analogues ($a< 0.65 $ au and $M<0.2 M_\oplus$) and 135 Mars analogues ($a> 1.3 $ au and $M<0.2 M_\oplus$) in systems with Venus and Earth analogues clearly identified. 

\section{Results}
\label{sec:results}

\begin{figure}
\centering
\includegraphics[width=\linewidth]{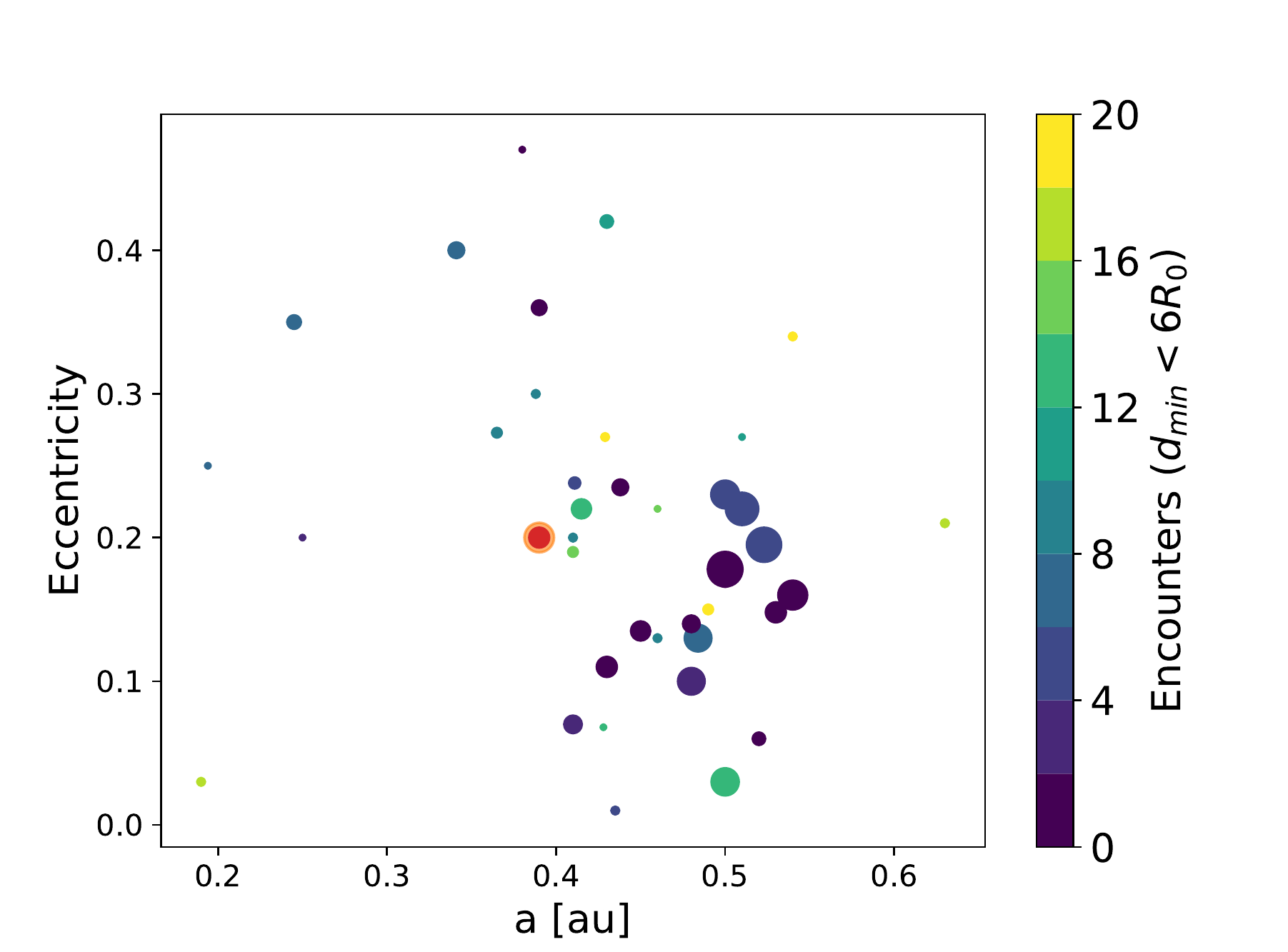}
\caption{Orbital parameters of formed Mercury analogues. The plot consists of 39 Mercury analogues with the Mercury planet indicated by the red circle (the enlarged orange circle indicates a hypothetical chondritic Mercury). The area of the circles scales linearly with the mass of the objects while the color represents their times of major encounters with proto-Venus whose $d_{min}<6R_0$  (see Figure \ref{fig:eg}).We note that Mercury analogues with masses larger than 0.3 Mercury mass have inclinations ($5^{\circ}-13^{\circ}$) close to present-day Mercury inclination, $7^{\circ}$. \label{fig:ae}}
\end{figure}

\cite{Hansen2009} has already noted the orbital overlaps between proto-Mercury and proto-Venus in the early stage of planetary accretion \cite[see Figure 6 of][]{Hansen2009}. However, no details about proto-Mercury and proto-Venus encounters are reported therein. All our formed Mercury analogues are scattered inwards by proto-Venus. This is not surprising as the potential well is deep in the inner solar system and thus violent encounters are necessary to shrink proto-Mercury's orbit. In figure \ref{fig:eg}, we show the encounter statistics between proto-Mercury and proto-Venus in a typical run. Both Mercury (green curve) and Venus (red curve) analogues accumulate materials quickly and are nearly fully-fledged at 40 Myr. Subsequently, a series of close encounters scatter proto-Mercury inside of proto-Venus. Here we focus on encounters after proto-Venus gains 0.7 Venus mass which can potentially remove proto-Mercury's mantle \citep{Deng2020}. We term these events as major encounters. Mass transfer in extreme close encounters can also happen between low mass embryos but ignored here because of the lack of available systematic studies. 

In the exemplary run of figure \ref{fig:eg}, major encounters have a relative velocity at infinity  $v_\infty \sim$ 10 km/s.  In major encounters, $v_\infty$ shows no correlation with the closest approach, $d_{min}$ which is typical of our simulations. The value of $v_\infty$ ranges from 2 km/s to 10 km/s for major encounters in our simulations. The frequency of major encounter scales with the square of $d_{min}$ when $d_{min}$ is sufficiently large simply reflecting the geometric cross section scaling. We note that all encounters with $d_{min}$ smaller than three Hill radii are recorded in our simulations. The occurrence rate of major encounter with $d_{min}$ smaller than 10$R_0$ (Venus radius, $R_0$) is rather irregular so that a statistical study is desirable.

\begin{figure}
\centering
\includegraphics[width=\linewidth]{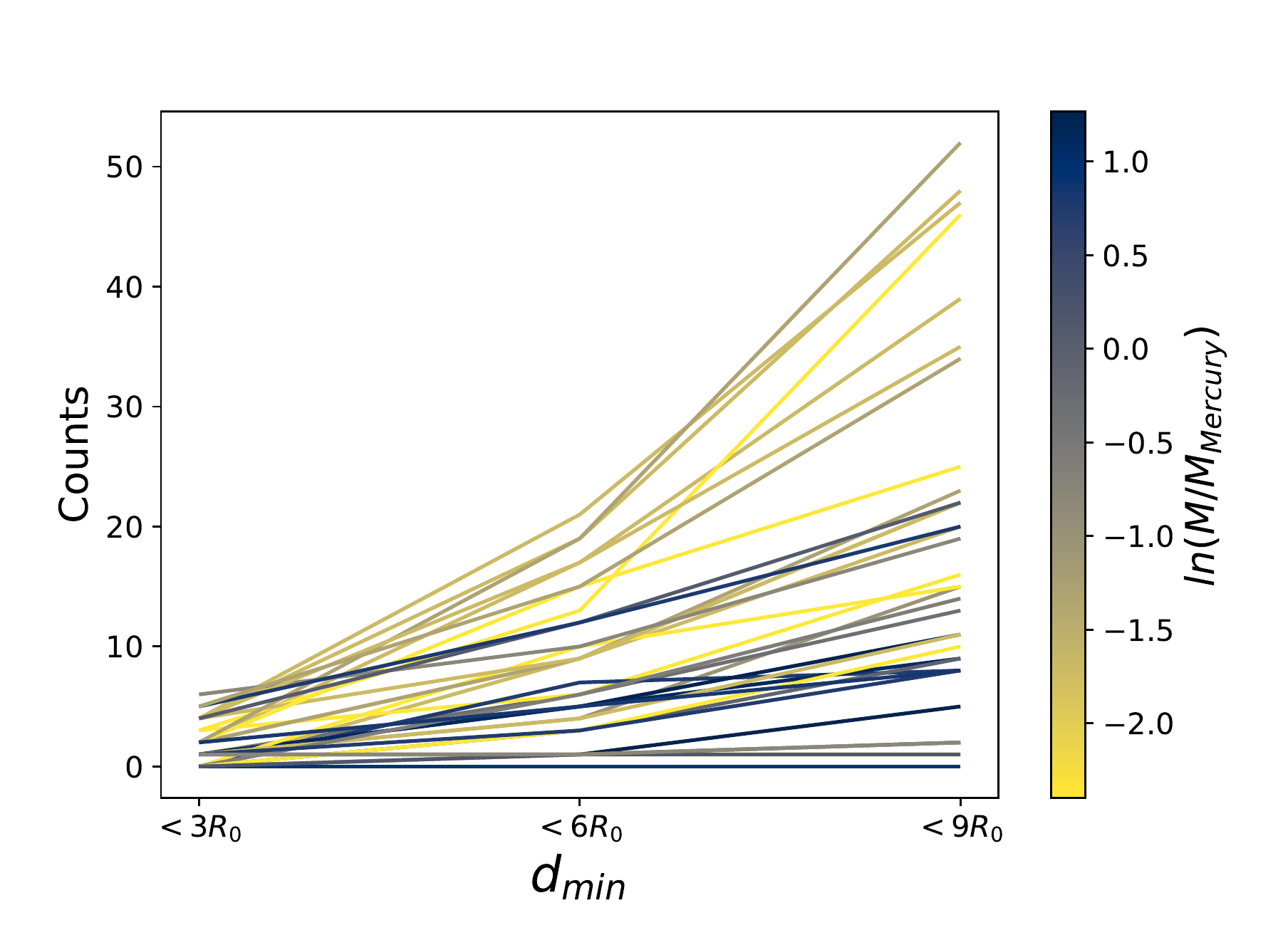}
\caption{The number of proto-Mercury and proto-Venus major encounters (see Figure \ref{fig:eg}) with $d_{min}<3R_0, 6R_0, 9R_0$ respectively. The line colors indicate the mass of the formed Mercury analogues in logarithmic scale. The median of encounter number for $d_{min}<3R_0, 6R_0, 9R_0$ are 1, 6, 13 respectively. \label{fig:mercury}}
\end{figure}

We plot in figure \ref{fig:ae} the orbital parameters of all the Mercury analogues. The small Mercury analogues show large scatter in both the semi-major axis and eccentricity. However, there is a clustering of 
Mercury analogues around 0.5 au and eccentricity = 0.2. This group of Mercury analogues are more massive than Mercury or even a chondritic Mercury (a hypothetical Mercury with iron silicate mass ratio equals to 3:7). They also lie too close to Venus analogues \citep[see also][Figure 2]{Clement2019}. However, they are in line with the tidal mantle stripping model of \cite{Deng2020} as mass transfer to Venus analogues (due to tidal interaction but ignored in N-body simulations) tends to shrink Mercury analogue's semi-major axis towards present-day Mercury value, 0.39 au. The remaining question is whether these encounters are frequent and violent enough to remove proto-Mercury's mantle significantly.

We calculated the number of extreme major encounters with $d_{min}$ smaller than certain values ($3R_0, 6R_0, 9R_0$). It is noteworthy that Venus is able to deflect the orbit of a chondritic Mercury by $~1^{\circ}$ when $d_{min}=9R_0$ and $v_\infty=10$ km/s in a test hydrodynamic simulation similar to those of \cite{Deng2020}.  In figure \ref{fig:mercury}, the lighter Mercury analogues experienced more extreme major encounters in general. Three Mercury analogues (0.27, 0.45, 2.18 Mercury mass) have more than 4  (5, 6, 5) major encounters with $d_{min}<3R_0$. We do not simply take them as successful realisations of the tidal mantle stripping hypothesis of \cite{Deng2020} (see section \ref{sec:diss}). The median of encounter number for $d_{min}<3R_0, 6R_0, 9R_0$ are 1, 6, 13 respectively. Extreme major encounters can dissipate orbital energy and even lead to mass transfer \citep{Deng2020}. Our encounter statistics necessitate proper treatment of tidal interaction in the study of Mercury formation. 

\begin{figure*}
\centering
\includegraphics[width=0.9\linewidth]{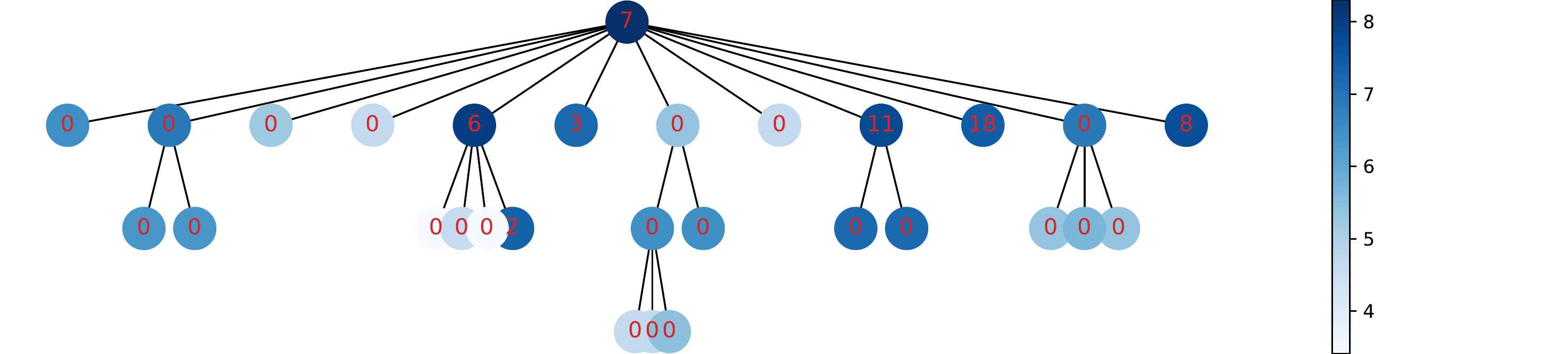}
\caption{The merger and encounter history of a Mercury analogue of 0.11 Earth mass. The number label the times of major encounter with proto-Venus whose $d_{min}<6R_0$ while the color represents the life time of embryos in logarithmic scale, i.e., log (life time in years).  \label{fig:tree}}
\end{figure*}

\begin{figure}
\centering
\includegraphics[width=\linewidth]{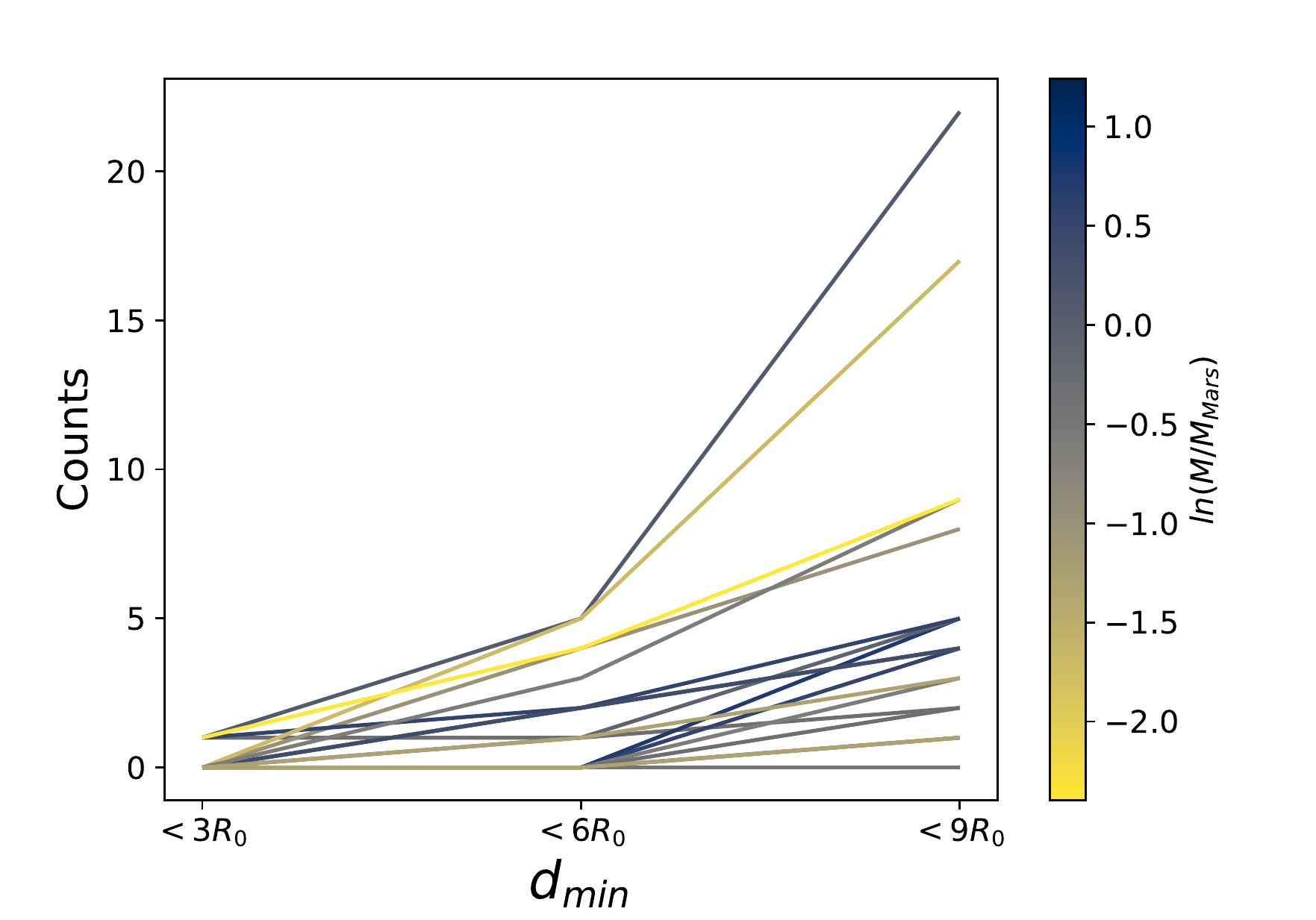}
\caption{The number of proto-Mars and proto-Earth major encounters similar to figure \ref{fig:mercury}. The median of encounter number for $d_{min}<3R_0, 6R_0, 9R_0$ are 0, 0, 1 respectively. \label{fig:mars}}
\end{figure}

We may wonder why Mars is not metal enriched as Mercury. Beyond 1 au, the potential well is shallow so moderate encounters can already scatter Mars from $\sim 1$ au to 1.5 au. We carried out similar major encounter analysis for 39 randomly chosen Mars analogues (the trend does not change when more samples are included). In figure \ref{fig:mars}, we find only about 1 major encounter at $d_{min}<9R_0$ between proto-Earth and proto-Mars. Most major encounters have $d_{min}$ tens of $R_0$ and $v_\infty\sim 5$ km/s. In some cases, Mars orbit can even diffuse outwards by indirectly exchanging angular momentum with proto-Earth through smaller objects \citep{Hansen2009}.

\section{Discussion}
\label{sec:diss}

We have shown clear evidence of extreme major encounters between proto-Mercury and proto-Venus in N-body simulations of terrestrial planet formation. We refrain from concluding on the probability of  proto-Mercury mantle striping \citep{Deng2020}. During planetary accretion, extreme encounters happen between Mercury-Venus pairs of different masses, $v_\infty$, $d_{min}$ and spin. The parameter space explored by \cite{Deng2020} is limited to $v_\infty<3$ km/s and $d_{min}<2R_0$ so results therein are not fully applicable here. A simple classification of planetary close encounter outcomes based on the Roche radius would be imprudent as \cite{Deng2020} clearly shows spins of planets make a huge different.

Aside from the limited knowledge in planetary encounters (due to the high dimensional parameter space), we have not considered all interactions between proto-Mercury and proto-Venus in our analysis above. In figure \ref{fig:eg}, we only focus on the two objects bearing the identity of the final Mercury and Venus analogues. Embryos merged into the final Mercury analogue may have experienced extreme encounter with proto-Venus as well. In figure \ref{fig:tree}, we demonstrate the growth and major encounter with proto-Venus for a Mercury analogue. Indeed, some embryos experience major encounters with proto-Venus and then merge into the final Mercury analogue. 
In this sense, the encounter times discussed above is a lower limit for the study of tidal mantle striping. 

On the other hand, the encounter statistics may not be accurate themselves. As we noted above, tidal dissipation at major encounters tends to bring proto-Mercury closer to proto-Venus \citep{Deng2020}. We expect more encounters at smaller $d_{min}$ if the dissipation caused by tidal interaction (wave excitation and/or mass transfer) is considered. Hybrid N-body and hydrodynamic simulations are necessary to address the formation of Mercury self-consistently. 

The low $v_\infty$ in Mercury-Venus encounters casts further doubt on the single hit-and-run giant impact scenario for Mercury formation which requires a typical impact velocity $v_{imp}\sim$20 km/s \citep{Asphaug2014,Chau2018}. Multiple impacts scenario at lower $v_{imp}$ may reconcile with the observation of moderate volatiles on Mercury more easily than a single impact \citep{Chau2018}. However, repeatedly hit-and-run collisions between proto-Venus and proto-Mercury are much less likely than repeatedly close encounters out of simple geometric cross section consideration. 

\section{Conclusion}
\label{sec:con}

We carried out 250 N-body simulations of terrestrial planet formation. We found frequent close encounters between proto-Mercury and proto-Venus after the latter gained 0.7 Venus mass. About 13 such encounters have closest approach smaller than 9 Venus radii.  These encounters scatters proto-Mercury to the innermost solar system. However, the formed Mercury analogues have a slightly larger semi-major axis (0.5 au) than present day Mercury value (0.39 au). We expect tidal orbital decay to lead to more violent encounters which eventually remove proto-Mercury's mantle significantly and bring it further inward.

\section*{Acknowledgement}
H.D. acknowledge support from the Swiss National Science Foundation via an early postdoctoral mobility fellowship. We are grateful to comments from Sean Raymond on an early version of this paper which lead to significant improvements. We thank the referee, John Chambers, for his report which greatly improved the clarity of the paper. 


\bibliographystyle{mnras}
\bibliography{references}


\bsp	
\label{lastpage}
\end{document}